\documentclass[sigconf]{acmart}
\AtBeginDocument{%
  }

\copyrightyear{2024}
\acmYear{2024}
\setcopyright{rightsretained}
\acmConference[RecSys '24]{18th ACM Conference on Recommender Systems}{October 14--18, 2024}{Bari, Italy}
\acmBooktitle{18th ACM Conference on Recommender Systems (RecSys '24), October 14--18, 2024, Bari, Italy}\acmDOI{10.1145/3640457.3688045}
\acmISBN{979-8-4007-0505-2/24/10}

\begin{document}

\title{Dynamic Product Image Generation and Recommendation at Scale for Personalized E-commerce}

\author{\'{A}d\'{a}m Tibor Czapp}
\email{adam-tibor.c@taboola.com}
\affiliation{%
  \institution{Taboola Budapest}
  \city{Budapest}
  \country{Hungary}
}
\author{M\'{a}ty\'{a}s Jani}
\email{matyas.j@taboola.com}
\affiliation{%
  \institution{Taboola Budapest}
  \city{Budapest}
  \country{Hungary}
}
\author{B\'{a}lint Domi\'{a}n}
\email{balint.d@taboola.com}
\affiliation{%
  \institution{Taboola Budapest}
  \city{Budapest}
  \country{Hungary}
}
\author{Bal\'{a}zs Hidasi}
\email{balazs.h@taboola.com}
\affiliation{%
  \institution{Taboola Budapest}
  \city{Budapest}
  \country{Hungary}
}


\begin{abstract}
Coupling latent diffusion based image generation with contextual bandits enables the creation of eye-catching personalized product images at scale that was previously either impossible or too expensive. In this paper we showcase how we utilized these technologies to increase user engagement with recommendations in online retargeting campaigns for e-commerce.
\end{abstract}



\maketitle

\section{Introduction}

The engagement of users with recommended products is greatly influenced by their presentation~\cite{goswami2011study}, which is second only to their relevance. This is especially true in online advertisement where the users' primary focus is not on the recommendations. Creatives of ad campaigns are often designed with great care, but this approach does not scale for product level ad campaigns (e.g.~retargeting, Dynamic Product Ads (DPA)) where each item of the product catalog is subject to be recommended on any of the ad placements with different aspect ratios. The common approach is to show the original product image, optionally with additional design elements. We improve upon this by using image generation methods and place the products in appropriate environments. These more eye-catching creatives increase user engagement. This solution is also useful for enhancing user generated product photos (e.g.~on marketplaces) that might have been taken in less appealing environments.

\section{Generating product images}
We designed a novel feature for our recommender system that creates eye-catching product images in the given size by generating the surrounding background. Background generation utilizes Stable Diffusion~\cite{stablediffusion} -- a popular diffusion~\cite{song2020score} based image generation model -- through the diffusers~\cite{von-platen-etal-2022-diffusers} package. The model is prompted with predefined prompts that describe environments appropriate for the product category. The product itself is not modified in any way.

\begin{figure}
  \centering
  \includegraphics[width=0.35\textwidth]{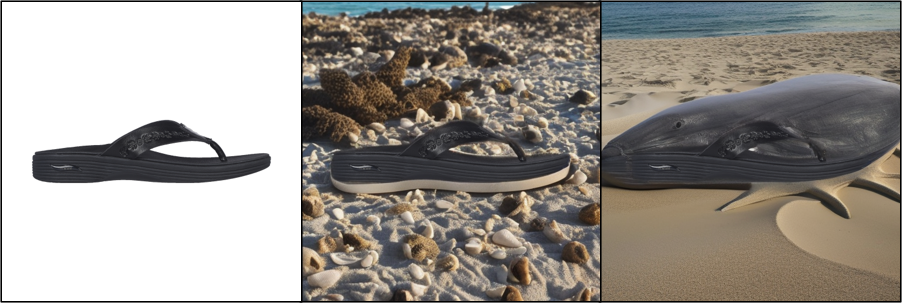}
  \Description[Examples for when inpainting extends the product by drawing to it.]{Left: original product image of a black slipper. Middle: the same slipper is on a rocky beach, but its sole is thicker, and its bottom part is white. Right: the slipper is on a sandy beach, and it is barely recognizable as inpainting extended it to an abstract object that slightly resembles a whale.}
  \caption{Mild and extreme artifacts produced by inpainting.}
  \label{fig:artifact}
\end{figure}

\begin{figure}
  \centering
  \includegraphics[width=0.47\textwidth]{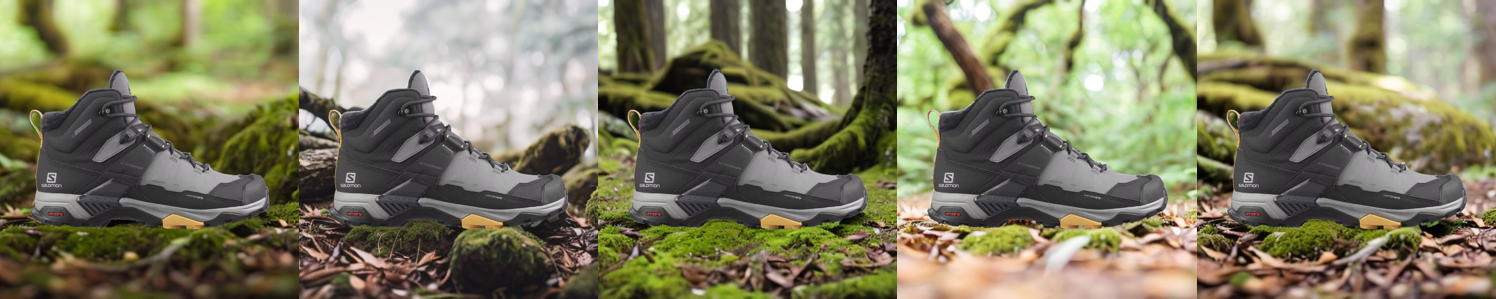}
  \Description[Examples of changing the color palette and the background composition of an image of a hiking shoe in the forest through conditioning.]{From left to right: (1) the lower part of the image is darker green with mossy rocks, the upper part is lighter green with leaves; (2) the lower part is predominantly brown with fallen leaves, to upper part is whitish with bright lights; (3) the dominant color in the whole image is bright green and there are trees in the background; (4) the lower part is light brown, the upper part is light green; (5) the lower part is dark brown and the upper part is green.}
  \caption{Examples of color variations through conditioning.}
  \label{fig:cond}
\end{figure}

\begin{figure}
  \centering
  \includegraphics[width=0.47\textwidth]{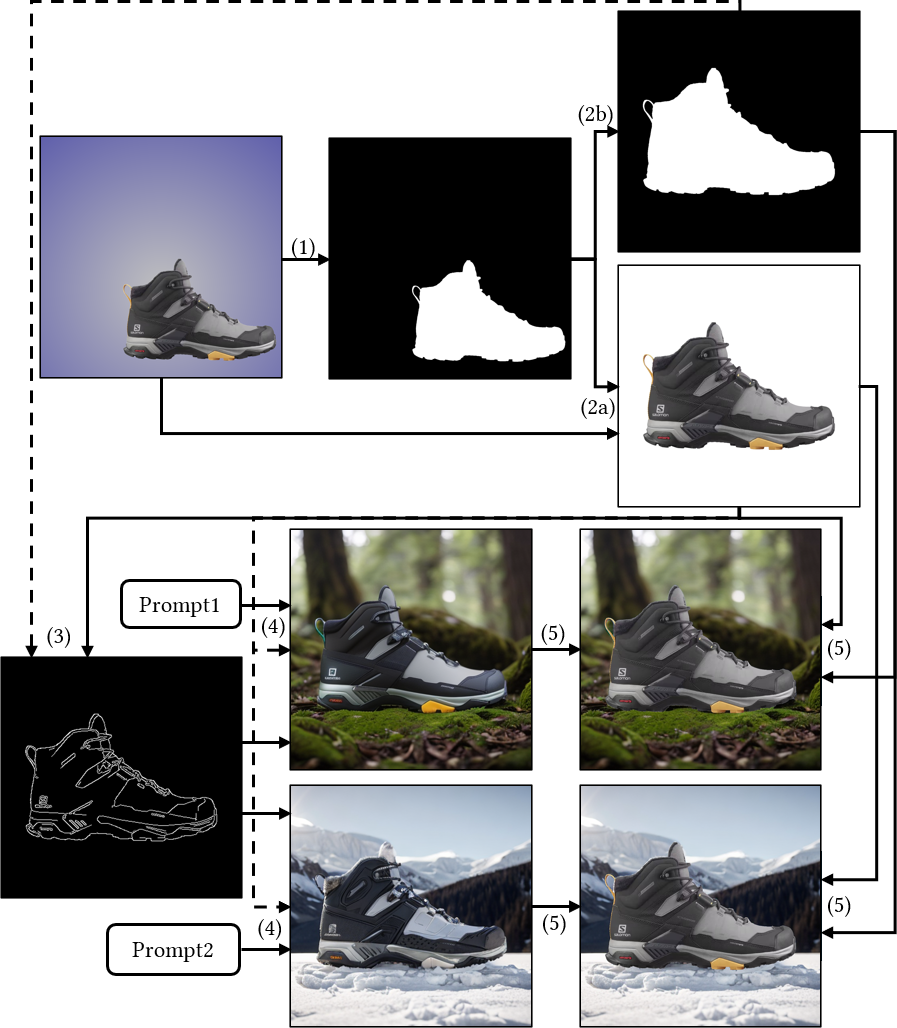}
  \Description[This figure depicts the pipeline of the image generation.]{This figure depicts the pipeline of the image generation.}
  \caption{Main steps of the background generation pipeline.}
  \label{fig:pipeline}
\end{figure}

\textbf{AI generated images in commercial applications:} Even with the rapid improvement of the technology, generative models -- by their nature -- are prone to create imperfect images. Fortunately, the application domain is somewhat permissive to smaller imperfections. Advertisements are rarely the main focus of the users, thus minor issues are likely overlooked and forgotten. But it is still crucial to build a pipeline where the chance of critical failures is low. The naive approach for this task is inpainting: mask the product and generate the background around it. However, this approach often produces artifacts by extending the product with virtual parts (see Figure~\ref{fig:artifact}). A better solution is to utilize ControlNet~\cite{controlnet}, which allows for the injection of additional constraints into the image generation process. Our pipeline uses the edges of the product as the constraint in ControlNet. While this approach does draw under the product mask, it creates an object similar to the product that is later replaced by the real product. This significantly reduces the artifacts and ``floating product'' images as well. We further improved this solution by conditioning the generation on the product. This technique reduces visible outlines and allows for subtly adapting the color palette and the composition without changing the prompt (see Figure~\ref{fig:cond}).

The main steps of the generation pipeline are shown on Figure~\ref{fig:pipeline}:
\begin{enumerate}
  \item Object detection and masking.
  \item Position \& scale the product (and mask) according to the placement. Remove original background, if applicable.
  \item Edge detection (optional: reinforce contours using the mask).
  \item Image generation using the edges as constraints (optional: also condition on the product to increase quality).
  \item Cut back the original product onto the generated image.
\end{enumerate}

\textbf{Production considerations:} The speed of image generation is too slow for the strict response time requirements of recommender systems, therefore the first request for a given (image, prompt, size) triplet only puts the request on a queue and the recommendation is served with the original image. Once the image is ready, the service calls back the recommender system that caches the generated image for future use. Costs are kept down by utilizing additional caches (e.g.~for masks) and grouping similar aspect ratios together. As no machine learning method is flawless, there are certain points where humans can intervene if necessary.

\textbf{Personalizing the experience:} Having a pretty background is not enough, it has to be appealing to the user that views it. Most products look good in multiple environments. Different users find different variations attractive and preferences might be influenced by where the recommendation is shown. Individually personalizing for every user--item--placement triplet is not possible due to the sparseness of the data. We deployed a solution similar to \cite{netflix_artwork} and use the LinUCB \cite{linucb1,linucb2} contextual bandit algorithm to select the prompt that has the best estimated CTR in the given context -- defined by user, item and placement features -- from the predefined prompt pool belonging to the product's category.


\begin{figure}
  \centering
  \includegraphics[width=0.47\textwidth]{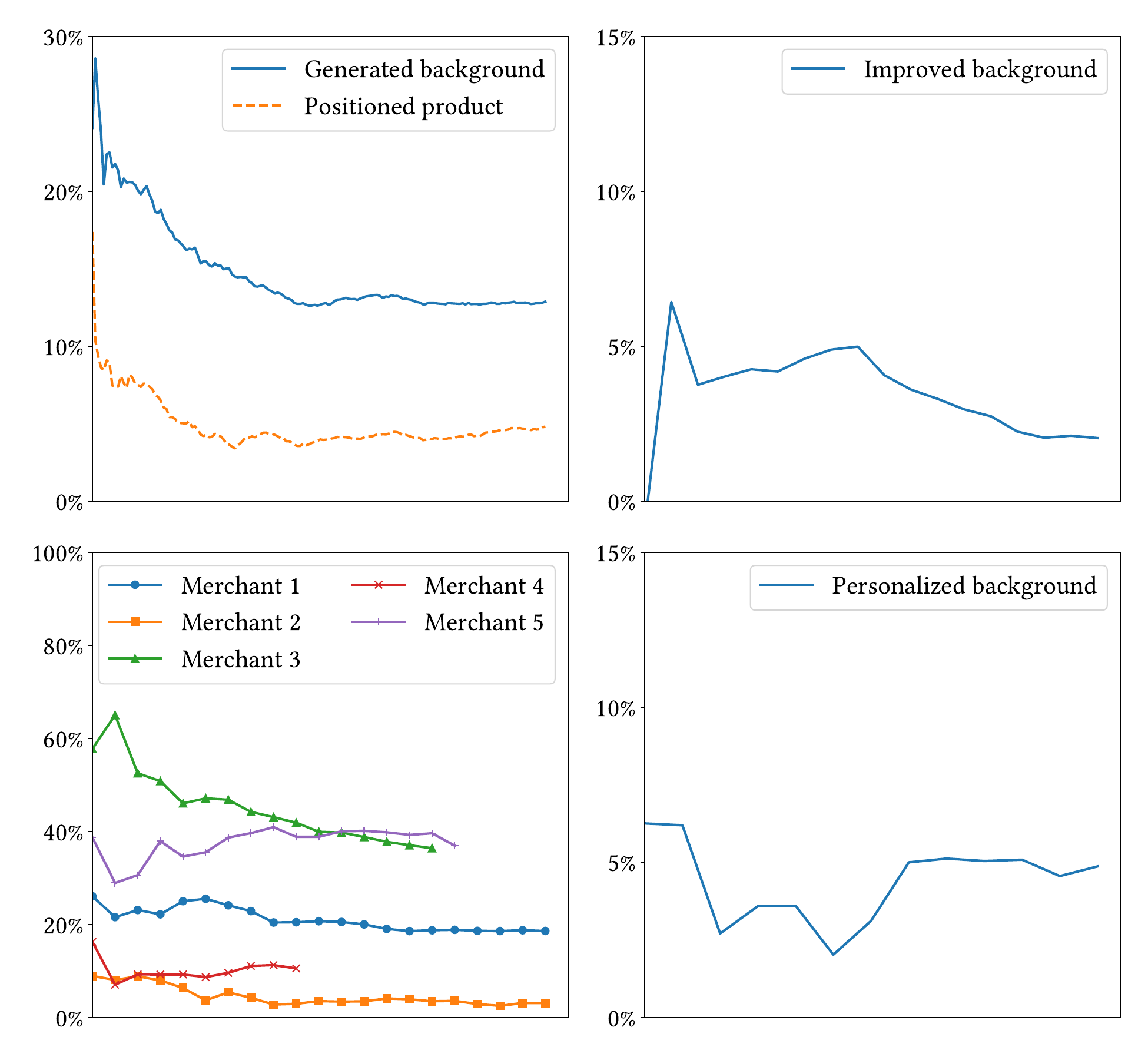}
  \Description[Results of the online A/B tests.]{Line plots depicting the results of the online A/B tests.}
  \caption{Relative CTR gains. (top left) phase I: positioned product on white \& generated background vs. original image; (top right) phase II: improved vs. old pipeline; (bottom left) phase II: generated image (improved pipeline) vs. original image; (bottom right) phase III: personalized vs. non-personalized background.}
  \label{fig:results}
\end{figure}

\section{Online tests}

Our approach was validated by online A/B tests. Multiple independent tests were performed over different timeframes and product catalogs. The sizes of the product catalogs were in the range of a few thousands to several tens of thousands items, and most of the products were in the apparel category (clothing, footwear, accessories, etc.). The primary metric was click-through rate (CTR), because creatives have direct effect on clicks only. However, we observed that other metrics (e.g.~number of conversions, cost per action (CPA)) also improved as an indirect effect of driving more users to the merchant's site. The test was conducted in three phases. Figure~\ref{fig:results} summarizes the results. All CTR gains are statistically significant at $p<0.05$.

\textbf{Phase I} focused on examining the added value of \textbf{generated backgrounds}, and was executed from September 2023 to February 2024. The performance of the generated images was compared to that of the original product images. Since the positioning and the size of the product can influence CTR, the output of step (2a) from Figure~\ref{fig:pipeline} was also included in the comparison as a separate group. The test validated both assumptions: (1) $\sim5\%$ improvement can be gained by simply paying attention to product positioning/scaling; (2) products with generated backgrounds performed even better with $\sim15\%$ gain over the baseline.

\textbf{Phase II} validated that the \textbf{improvements we made on the pipeline} resulted in additional CTR gains. This phase was also used to check the performance of the improved pipeline against the original product images \textbf{in multiple experiments}. While generated images always outperform the baseline, the relative CTR gain varies in a wide range ($\sim4-40\%$ in these experiments). The exact number mainly depends on the product catalog of the merchant, the ad placements, and the composition and quality of the original product images.

\textbf{Phase III} investigated the added benefit of \textbf{personalizing the backgrounds}. Three appropriate prompts were defined for each product category. In the treatment group, prompts are selected by the LinUCB algorithm based on context and user features, while the control group is served with images based on randomly selected prompts. Results suggest that even this lightweight personalization can further improve the performance of the system by $\sim5\%$.


\section*{Author bios}
\textbf{\'{A}d\'{a}m Tibor Czapp} and \textbf{B\'{a}lint Domi\'{a}n} are Machine Learning Engineers working on recommender systems \& algorithms. \textbf{M\'{a}ty\'{a}s Jani} is a Senior Machine Learning Software Engineer with 5+ years of experience in putting research results into live production systems, designing and implementing data pipelines and services around algorithms. \textbf{Bal\'{a}zs Hidasi} is a Leading Research Scientist with 15+ years of experience in machine learning and 10+ in leading machine learning and data science teams; he conducts research, directs research projects and oversees the machine learning related initiatives of Taboola Budapest (formerly Gravity R\&D).


\bibliographystyle{ACM-Reference-Format}
\bibliography{bggen}

\end{document}